\documentclass[a4paper,11pt]{article}
\usepackage[dvips]{graphicx}
\usepackage{amssymb}
\usepackage{amsmath}

\begin{document}

\title{The Shape of Compact Toroidal Dimensions $T^d_{\theta}$ and the Casimir Effect on $M^D\times T^d_{\theta}$ spacetime}
\author{V.K.Oikonomou\thanks{
voiko@physics.auth.gr}\\
Dept. of Theoretical Physics Aristotle University of Thessaloniki,\\
Thessaloniki 541 24 Greece\\
and\\
T.E.I. Serres} \maketitle

\begin{abstract}

We study the influence of the shape of compact dimensions to the
Casimir energy and Casimir force of a scalar field. We examine
both the massive and the massless scalar field. The total
spacetime topology is $M^D\times T^2_{\theta}$, where $M^D$ is the
$D$ dimensional Minkowski spacetime and $T^2_{\theta}$ the twisted
torus described by $R_1$, $R_2$ and $\theta$. For the case
$R_1=R_2$ we found that the massive bulk scalar field Casimir
energy is singular for $D$=even and this singularity is
$R$-dependent and remains even when the force is calculated. Also
the massless Casimir energy and force is regular only for $D=4$
(!). This is very interesting phenomenologically. We examine the
energy and force as a function of $\theta$. Also we address the
stabilization problem of the compact space. We also briefly
discuss some phenomenological implications.
\end{abstract}

\smallskip

\section*{Introduction}

The Casimir effect is one of the many macroscopic manifestations
of quantum fluctuations. Since the original paper of H. Casimir
the computation of the Casimir energy and Casimir force has
developed to a research area on its own, with many theoretical and
experimental applications \cite{Bordagreview}. The applications
are vast, varying from the calculation of the vacuum energy
between plates to cosmological implications. In most cases the
Casimir energy is affected from the geometry and topology of the
spacetime.

\noindent Many studies have focused on the calculation of the
Casimir energy in the presence of compactified space dimensions,
see for example \cite{elizalde}. The main interest is focused on
the sign of the Casimir energy and Casimir force. Concerning the
stabilization of the compact dimensions, there exist many
approaches in these issues. Some of these deal with the
stabilization of the compact extra dimensions through the radion
field. When the radion field acquires positive mass square due to
a repulsive (positive) Casimir force. The last requires the
presence of negative tension brane \cite{kanti}. A negative mass
square is due to attractive Casimir force and corresponds to
unstable minimum of the radion \cite{kanti}. Also calculations
have been performed in the presence of compact non-commutative
extra dimensions. The interest on these calculations is focused
mainly on the calculation of the Casimir force with respect to the
compact space. A negative Casimir energy is a very good feature of
these theories since it leads to a shrinking of the compact
dimensions. Also in some cases one loop corrections lead to
stabilization of the compact space. We shall discuss on this more
in the following. In addition, calculation of the Casimir energy
poses restrictions to the size of the extra dimensions, see for
example \cite{perivolaropoulos,Miltonnew}. However in most studies
where compact dimensions are taken into account, the main interest
concerning the compact space is focused on the volume of the extra
dimensions. With volume we mean the size of the radii of the
compact dimensions. Less interest has been given on the shape of
the extra dimensions. In the papers \cite{dienes} of K. Dienes,
the study was focused on the effect of the shape of the compact
space in the phenomenology of four dimensional spacetime. Also in
the paper of K. Kirsten and E. Elizalde \cite{elizaldekirsten} the
calculation of the Casimir energy for an arbitrary shaped two
dimensional toroidal surface was firstly performed. The results
found by K. Dienes are very interesting. Specifically it seems
that the shape of extra dimensions induce level crossings and
varying mass gaps, the elimination of light KK states and the
fascinating possibility of the alteration of the experimental
constraints for the extra dimensions. Also the ''shadowing''
process is a very interesting feature. K. Dienes deals with a
torus with twisted lattice, which we refer here as twisted torus
and we denote it $T^2_{\theta}$ for brevity. The twisted torus can
be seen in Fig.\ref{twistedtorus1}. The parameters describing the
twisted torus are $R_1$, $R_2$ and $\theta$.

\noindent In this article we shall include the effect of the shape
of compact dimensions to the Casimir energy. Our aim is to study
the Casimir energy and Casimir force for spacetime topologies
$M^D\times T^2_{\theta}$, with $M^D$ Minkowski space times and
$T^2_{\theta}$ the arbitrary shaped twisted two dimensional torus.
The study will be for the scalar field quantized with this
topology, both with and without mass. We are interested to see how
the Casimir energy and the corresponding Casimir force behaves as
a function of the shape of the extra dimensions parameter (which
is $\theta$, see below). This study will include the Casimir force
sign and we check if the form of the Casimir energy leads to a
stabilization of the compact space. Additionally we shall focus
where applicable to our four dimensional spacetime, which is the
most interesting case phenomenologically.

\noindent In section 1 we describe briefly the eigenfunctions and
eigenvalues of the scalar field for the $M^D\times T^2_{\theta}$
topology, following \cite{dienes}. In section 2 we compute the
Casimir energy and force for a massive scalar on $M^D\times
T^2_{\theta}$. In section 3 we do the same for the massless
scalar. The conclusions with a discussion follow in section 4.

\section{Eigenfunctions and Eigenvalues on $M^{D}\times T^2_{\theta}$}

We describe here the twisted torus $T^2_{\theta}$ and the
eigenfunctions and eigenvalues of the scalar field for the
spacetime $M^{D}\times T^2_{\theta}$. We shall follow the
presentation of K. Dienes \cite{dienes}.
\begin{figure}[h]
\begin{center}
\includegraphics[scale=0.6]{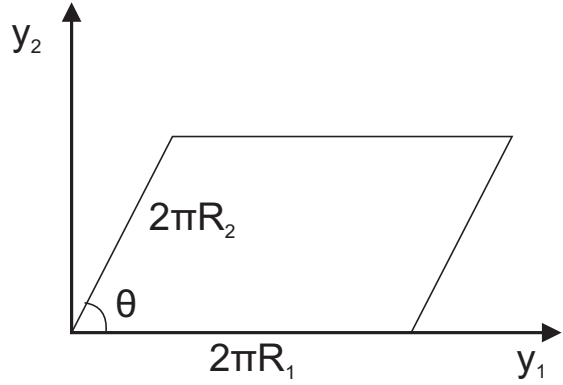}
\end{center}
\caption{The twisted torus described by $R_1$, $R_2$ and
$\theta$.} \label{twistedtorus1}
\end{figure}

\noindent The general twisted torus case described by Figure
\ref{twistedtorus1} with the torus radii $R_1$ and $R_2$ and also
with lattice angle $\theta$. The twisted torus $T^2_{\theta}$ is
realized as the flat space $R^2$ described by the coordinates
$y_1$ and $y_2$ with the identifications,
\begin{align}\label{ident}
&y_1\rightarrow y_1+2\pi R_1 \\& \notag y_2\rightarrow y_2\\&
\notag y_1\rightarrow y_1+2\pi R_2\cos\theta \\& \notag
y_2\rightarrow y_2+2\pi R_2\sin\theta
\end{align}
It is obvious that the value $\theta=\frac{\pi}{2}$ gives the
known torus $T^2$. Our aim here is to present the solutions to the
Laplace equation,
\begin{equation}
-\Delta_{ _{M^{D}\times T^2_{\theta}}}\phi(x,y_1,y_2)=\omega_{
_{M^{D}\times T^2_{\theta}}}\phi(x,y_1,y_2)
\end{equation}
for the scalar field in the spacetime $M^{D}\times T^2_{\theta}$
and also the eigenvalues (where $x$ denotes the space coordinates
of the $M^D$ spacetime coordinate). Demanding invariance under the
torus identifications (\ref{ident}), the eigenfunctions are,
\begin{equation}\label{harmexpans}
\phi(x,y_1,y_2)=\int
\frac{\mathrm{d}^{D-1}p}{(2\pi)^{D-1}}\sum_{n,n_1=-\infty}^{\infty}e^{i\vec{p}\vec{x}}e^{i\frac{n}{R_1}\big{(}y_1-\frac{y_2}{\tan
\theta}\big{)}+i\frac{n_1{\,}y_2}{R_2 \sin \theta}}.
\end{equation}
Now it follows that the eigenvalues for the massive scalar are,
\begin{equation}\label{eigenvalues}
\omega_{ _{M^{D}\times
T^2_{\theta}}}=\sum_{k=1}^{D-1}p_k^2+\frac{n^2}{\sin^2\theta
R_1^2}+\frac{n_1^2}{\sin^2\theta
R_2^2}-2\frac{nn_1}{R_1R_2}\cos\theta+m^2,
\end{equation}
and for the massless,
\begin{equation}\label{eigenvalues1}
\omega_{ _{M^{D}\times
T^2_{\theta}}}=\sum_{k=1}^{D-1}p_k^2+\frac{n^2}{\sin^2\theta
R_1^2}+\frac{n_1^2}{\sin^2\theta
R_2^2}-2\frac{nn_1}{R_1R_2}\cos\theta.
\end{equation}
The $\theta$ values are restricted to the range $0<\theta \leq
\frac{\pi}{2}$ without loss of generality. We shall use the
following notation in the next sections, namely,
\begin{align}\label{notation1}
&a=\frac{1}{\sin^2\theta R_1^2}\\&\notag c=\frac{1}{\sin^2\theta
R_2^2}\\&\notag b=-2\frac{1}{R_1R_2}\cos\theta
\end{align}
and also,
\begin{equation}\label{notation12}
\Delta=4ac-b^2=\frac{4}{R_1^2R_2^2\sin^2\theta}\big{(}1-\sin^2\theta\cos^2\theta\big{)}
\end{equation}
For later use note that $\Delta \geq 0$ for all $\theta$ values.
Our interest is mainly for the case $R_1=R_2$. Thus we shall try
to find how the changes of $\theta$ alter the Casimir energy.
However in the case with equal torus radii the shadowing
phenomenon is absent. We shall discuss on these issues in the
conclusions.

\section{Casimir energy and Casimir force for the massive scalar field on $M^D\times T^2_{\theta}$}

\subsection{Casimir energy with general $R_1$, $R_2$ and $\theta$}

We now calculate the Casimir energy for the massive scalar in the
$M^D\times T^2_{\theta}$ spacetime. Using relation
(\ref{eigenvalues}) and the notation of relations
(\ref{notation1}) and (\ref{notation12}), the Casimir energy for
the twisted torus reads (at the end we put $s=-\frac{1}{2}$),
\begin{align}\label{vlad}
\mathcal{E}_c(s)=\frac{1}{(2\pi)^{D-1}}\int
\mathrm{d}^{D-1}p\sum^{\infty}_{n,{\,}n_1=-\infty}\Big{[}\sum_{k=1}^{D-1}p_k^2+an^2+b{\,}nn_1+cn_1^2+m^2\Big{]}^{-s}.
\end{align}
In the end we put $s=-\frac{1}{2}$. Upon integrating over the
continuous dimensions using,
\begin{equation}\label{feynman}
\int
\mathrm{d}k^{D-1}\frac{1}{(k^2+A)^s}=\pi^{\frac{D-1}{2}}\frac{\Gamma(s-\frac{D-1}{2})}{\Gamma(s)}\frac{1}{A^{s-\frac{D-1}{2}}}
\end{equation}
relation (\ref{vlad}) becomes,
\begin{align}\label{pordoulis}
\mathcal{E}_c(s,a)=\frac{1}{(2\pi)^{D-1}}\pi^{\frac{D-1}{2}}\frac{\Gamma(s-\frac{D-1}{2})}{\Gamma(s)}\sum^{\infty}_{n,{\,}n_1=-\infty}\Big{[}an^2+b{\,}nn_1+cn_1^2+m^2\Big{]}^{\frac{D-1}{2}-s}.
\end{align}
Using the inhomogeneous Epstein zeta-like function
\cite{elizaldeoriginal,elizalde,elizaldekirsten,kirsten12,kirsten14},
\begin{equation}\label{inhomoge}
E(s;a,b,c;q)=\sum^{\infty
'}_{n,{\,}n_1=-\infty}\Big{[}an^2+b{\,}nn_1+cn_1^2+m^2\Big{]}^{-s}
\end{equation}
relation (\ref{pordoulis}) is written,
\begin{align}\label{pordoulis}
&\mathcal{E}_c(s,a)=\frac{1}{(2\pi)^{D-1}}\pi^{\frac{D-1}{2}}\frac{\Gamma(s-\frac{D-1}{2})}{\Gamma(s)}{\,}{\,}E(s-\frac{D-1}{2};a,b,c;m^2)\\&\notag
+\frac{1}{(2\pi)^{D-1}}\pi^{\frac{D-1}{2}}\frac{\Gamma(s-\frac{D-1}{2})}{\Gamma(s)}{\,}(m^2)^{-(s-\frac{D-1}{2})}.
\end{align}
Now the inhomogeneous can be expanded according to the extended
Chowla-Selberg formula \cite{elizaldeoriginal,elizaldekirsten},
\begin{align}\label{inchowlaselberg}
&E(s;a,b,c;q)=2{\,}\zeta_{EH}(s,\frac{q}{a}){\,}a^{-s}+\frac{2^{2s}\sqrt{\pi}{\,}a^{s-1}}{\Gamma(s)\Delta^{s-\frac{1}{2}}}\Gamma(s-\frac{1}{2})\zeta_{EH}(s-\frac{1}{2},\frac{4aq}{\Delta})
\\&\notag
+\frac{2^{s+\frac{5}{2}}{\,}\pi^s}{\Gamma(s)\sqrt{a}}\sum_{n=1}^{\infty}n^{s-\frac{1}{2}}\cos(\frac{n\pi
b}{a})\sum_{d/n}d^{1-2s}\Big{(}\Delta+\frac{4aq}{d^2}\Big{)}^{-\frac{s}{2}+\frac{1}{4}}K_{s-\frac{1}{2}}\Big{(}\frac{\pi
n}{a}\sqrt{\Delta+\frac{4aq}{d^2}}{\,}\Big{)}
\end{align}
which is defined for $\Delta \geq 0$ which in our case holds as we
saw previously. In the above $\zeta_{EH}(s;p)$ stands for the
inhomogeneous Epstein zeta
\cite{elizalde,elizaldeoriginal,elizaldekirsten,kirstenbook,kirsten12,kirsten14},
\begin{align}\label{gone}
\zeta_{EH}(s;p)&=\frac{1}{2}\sum_{n=-\infty}^{\infty
'}\big{(}n^2+p\big{)}^{-s}
\\& \notag =-\frac{p^{-s}}{2}+\frac{\sqrt{\pi}{\,}{\,}\Gamma(s-\frac{1}{2})}{2\Gamma(s)}{\,}p^{-s+\frac{1}{2}}+\frac{2\pi^s{\,}p^{-s+\frac{1}{2}}}{\Gamma(s)}\sum_{n=1}^{\infty}n^{s-\frac{1}{2}}K_{s-\frac{1}{2}}\big{(}2\pi
n\sqrt{p}\big{)}
\end{align}

\begin{figure}[h]
\begin{center}
\includegraphics[scale=1.0]{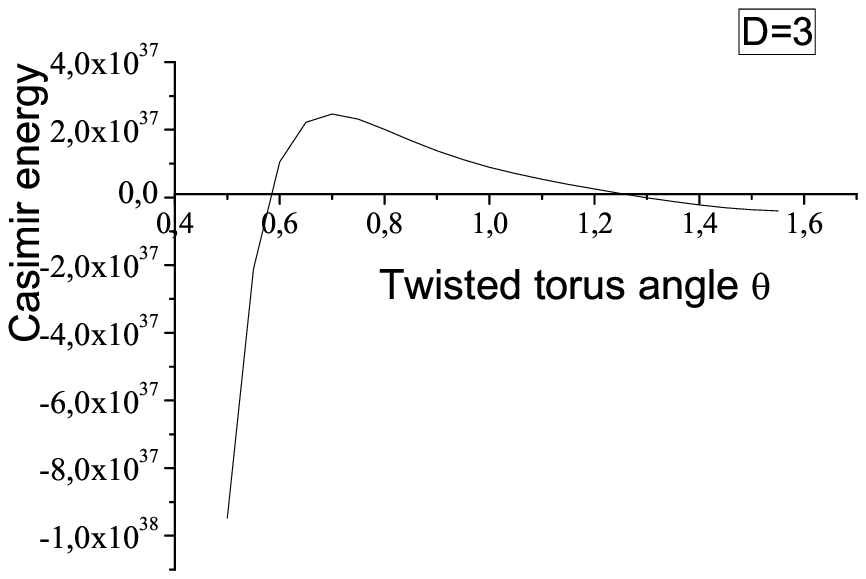}
\end{center}
\caption{The Casimir energy $\mathcal{E}_{c}$ as a function of
$\theta$, the twisted torus angle, for $D=3$. (Massive case)}
\label{energymassive1}
\end{figure}

\begin{figure}[h]
\begin{center}
\includegraphics[scale=1.0]{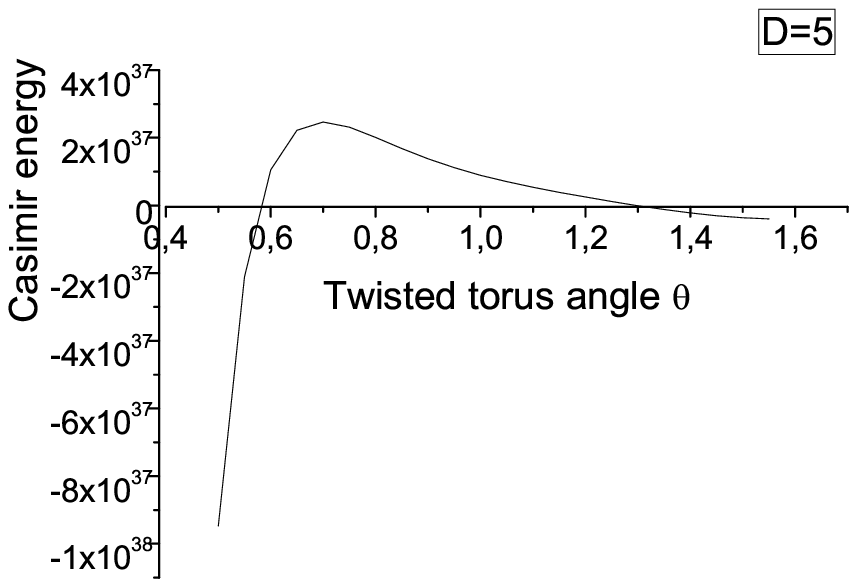}
\end{center}
\caption{The Casimir energy $\mathcal{E}_c$ as a function of
$\theta$, the twisted torus angle, for $D=5$. (Massive case)}
\label{energymassive2}
\end{figure}

Thus with the help of (\ref{inchowlaselberg}), the Casimir energy
of (\ref{pordoulis}) can be written,
\begin{align}\label{mpouxesas}
&\mathcal{E}_c(s)=\frac{1}{(2\pi)^{D-1}}\pi^{\frac{D-1}{2}}\frac{\Gamma(s-\frac{D-1}{2})}{\Gamma(s)}{\,}{\,}\times\\&
\notag
\Bigg{(}2{\,}\zeta_{EH}(s-\frac{D-1}{2},\frac{q}{a}){\,}a^{-(s-\frac{D-1}{2})}+\frac{2^{2(s-\frac{D-1}{2})}\sqrt{\pi}{\,}a^{s-\frac{D-1}{2}-1}}{\Gamma(s-\frac{D-1}{2}){\,}\Delta^{s-\frac{D-1}{2}-\frac{1}{2}}}\times\\&\notag\Gamma(s-\frac{D-1}{2}-\frac{1}{2}){\,}{\,}\zeta_{EH}(s-\frac{D-1}{2}-\frac{1}{2},\frac{4aq}{\Delta})
\\&\notag
+\frac{2^{s-\frac{D-1}{2}+\frac{5}{2}}{\,}\pi^{s-\frac{D-1}{2}}}{\Gamma(s-\frac{D-1}{2})\sqrt{a}}\sum_{n=1}^{\infty}n^{s-\frac{D-1}{2}-\frac{1}{2}}\cos(\frac{n\pi
b}{a})\times \\&\notag
\sum_{d/n}d^{1-2(s-\frac{D-1}{2})}\Big{(}\Delta+\frac{4aq}{d^2}\Big{)}^{-\frac{s-\frac{D-1}{2}}{2}+\frac{1}{4}}K_{s-\frac{D-1}{2}-\frac{1}{2}}\Big{(}\frac{\pi
n}{a}\sqrt{\Delta+\frac{4aq}{d^2}}\Big{)}\Bigg{)}
\\&\notag
+\frac{1}{(2\pi)^{D-1}}\pi^{\frac{D-1}{2}}\frac{\Gamma(s-\frac{D-1}{2})}{\Gamma(s)}{\,}(m^2)^{-(s-\frac{D-1}{2})}.
\end{align}
Now let us write each term of relation (\ref{mpouxesas})
separately, in order to present the details of the calculations.
The first term is written,
\begin{align}\label{protos}
&2{\,}\zeta_{EH}(s-\frac{D-1}{2},\frac{q}{a}){\,}a^{-(s-\frac{D-1}{2})}=-(\frac{q}{a})^{-(s-\frac{D-1}{2})}a^{-(s-\frac{D-1}{2})}\\&\notag
+a^{-(s-\frac{D-1}{2})}\frac{\sqrt{\pi}{\,}{\,}\Gamma(s-\frac{D-1}{2}-\frac{1}{2})}{\Gamma(s-\frac{D-1}{2})}{\,}(\frac{q}{a})^{-(s-\frac{D-1}{2})+\frac{1}{2}}\\&\notag
+a^{-(s-\frac{D-1}{2})}(\frac{q}{a})^{-\frac{1}{2}(s-\frac{D-1}{2})+\frac{1}{4}}\frac{2{\,}\pi^{s-\frac{D-1}{2}}}{\Gamma(s-\frac{D-1}{2})}\sum_{n=1}^{\infty}n^{s-\frac{D-1}{2}-\frac{1}{2}}K_{s-\frac{D-1}{2}-\frac{1}{2}}\big{(}2\pi
n\sqrt{\frac{q}{a}}\big{)}
\end{align}
The second term is written,
\begin{align}\label{deyteros}
&\frac{2^{2(s-\frac{D-1}{2})}\sqrt{\pi}{\,}a^{s-\frac{D-1}{2}-1}}{\Gamma(s-\frac{D-1}{2}){\,}\Delta^{s-\frac{D-1}{2}-\frac{1}{2}}}\Gamma(s-\frac{D-1}{2}-\frac{1}{2}){\,}{\,}\zeta_{EH}(s-\frac{D-1}{2}-\frac{1}{2},\frac{4aq}{\Delta})=
\\&\notag-\frac{2^{2(s-\frac{D-1}{2})}\sqrt{\pi}{\,}a^{s-\frac{D-1}{2}-1}\Gamma(s-\frac{D-1}{2}-\frac{1}{2})}{2{\,}\Gamma(s-\frac{D-1}{2}){\,}\Delta^{s-\frac{D-1}{2}-\frac{1}{2}}}\big{(}\frac{4aq}{\Delta}\big{)}^{-(s-\frac{D-1}{2}-\frac{1}{2})}
\\&\notag+\frac{2^{2(s-\frac{D-1}{2})}\pi{\,}a^{s-\frac{D-1}{2}-1}\Gamma(s-\frac{D-1}{2}-1)}{4{\,}\Gamma(s-\frac{D-1}{2}){\,}\Delta^{s-\frac{D-1}{2}-\frac{1}{2}}}\big{(}\frac{4aq}{\Delta}\big{)}^{-(s-\frac{D-1}{2}-\frac{1}{2})+\frac{1}{2}}
\\&\notag+\frac{2^{2(s-\frac{D-1}{2})}\sqrt{\pi}{\,}a^{s-\frac{D-1}{2}-1}{\,}\pi^{s-\frac{D-1}{2}-\frac{1}{2}}}{\Gamma(s-\frac{D-1}{2}){\,}\Delta^{s-\frac{D-1}{2}-\frac{1}{2}}}\big{(}\frac{4aq}{\Delta}\big{)}^{-(s-\frac{D-1}{2}-\frac{1}{2})+\frac{1}{4}}\times
\\&\notag \sum_{n=1}^{\infty}n^{s-\frac{D-1}{2}-1}K_{s-\frac{D-1}{2}-\frac{1}{2}}\Big{(}2\pi
n\sqrt{\frac{4aq}{\Delta}}{\,}\Big{)}
\end{align}
From the above two relations, (\ref{protos}) and (\ref{deyteros})
it is easily seen that various cancellations occur, for example
the first term of the second term cancels the second term of the
first term. After some algebra we finally obtain,
\begin{align}\label{teliki1}
&\mathcal{E}_c(s)=\frac{1}{(2\pi)^{D-1}}\frac{\pi^{\frac{D-1}{2}}}{\Gamma(s)}{\,}{\,}\times\\&
\notag
\Bigg{(}2{\,}\pi^{s-\frac{D-1}{2}}{\,}a^{\frac{1}{2}(s-\frac{D-1}{2})+\frac{1}{4}}{\,}q^{-\frac{1}{2}(s-\frac{D-1}{2})+\frac{1}{4}}\sum_{n=1}^{\infty}n^{s-\frac{D-1}{2}-\frac{1}{2}}K_{s-\frac{D-1}{2}-\frac{1}{2}}\big{(}2\pi
n\sqrt{\frac{q}{a}}\big{)}
\\&\notag+\frac{2^{2(s-\frac{D-1}{2})}\pi{\,}a^{s-\frac{D-1}{2}-1}\Gamma(s-\frac{D-1}{2}-1)}{4{\,}\Delta^{s-\frac{D-1}{2}-\frac{1}{2}}}\big{(}\frac{4aq}{\Delta}\big{)}^{-(s-\frac{D-1}{2}-\frac{1}{2})+\frac{1}{2}}
\\&\notag+\frac{2^{2(s-\frac{D-1}{2})}\sqrt{\pi}{\,}a^{s-\frac{D-1}{2}-1}{\,}\pi^{s-\frac{D-1}{2}-\frac{1}{2}}}{\Delta^{s-\frac{D-1}{2}-\frac{1}{2}}}\big{(}\frac{4aq}{\Delta}\big{)}^{-(s-\frac{D-1}{2}-\frac{1}{2})+\frac{1}{4}}\times
\\&\notag \sum_{n=1}^{\infty}n^{s-\frac{D-1}{2}-1}K_{s-\frac{D-1}{2}-\frac{1}{2}}\big{(}2\pi
n\sqrt{\frac{4aq}{\Delta}}\big{)}
\\&\notag
+\frac{2^{s-\frac{D-1}{2}+\frac{5}{2}}{\,}\pi^{s-\frac{D-1}{2}}}{\sqrt{a}}\sum_{n=1}^{\infty}n^{s-\frac{D-1}{2}-\frac{1}{2}}\cos(\frac{n\pi
b}{a})\times \\&\notag
\sum_{d/n}d^{1-2(s-\frac{D-1}{2})}\Big{(}\Delta+\frac{4aq}{d^2}\Big{)}^{-\frac{s-\frac{D-1}{2}}{2}+\frac{1}{4}}K_{s-\frac{D-1}{2}-\frac{1}{2}}\Big{(}\frac{\pi
n}{a}\sqrt{\Delta+\frac{4aq}{d^2}}{\,}\Big{)}\Bigg{)}
\end{align}

\begin{figure}[h]
\begin{center}
\includegraphics[scale=.8]{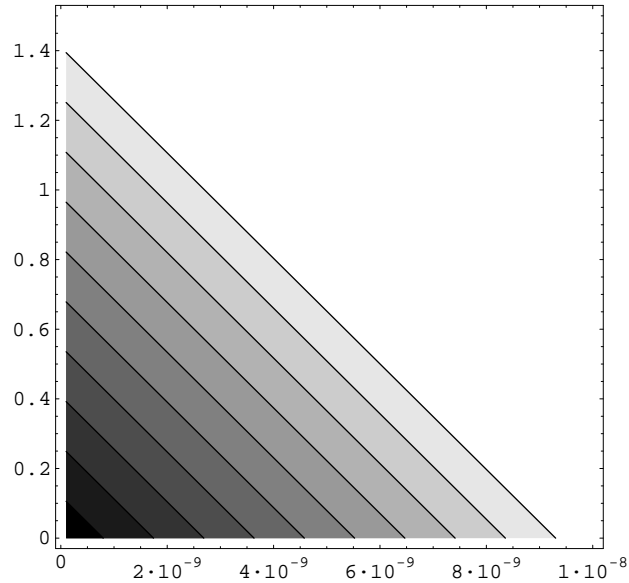}
\end{center}
\caption{Contour plot of the Casimir energy $\mathcal{E}_c$ as a
function of $\theta$ (vertical), the twisted torus angle, and $R$,
the compact radius (horizontal), for $D=3$. (Massive case)}
\label{energymassive3}
\end{figure}
\noindent We can easily see that relation (\ref{teliki1}) contains
a singularity when $D$=even. Indeed the singularity is due to the
gamma function $\Gamma(s-\frac{D-1}{2}-1)$. Thus our four
dimensional spacetime is excluded from the study. However we will
continue to present the results because these are very interesting
mathematically and also in order to have a clear picture for all
cases.

\subsection{The case $R_1=R_2$}

Now we specify our result (\ref{teliki1}) to the case $R_1=R_2=R$.
Thus the only parameter that characterizes the shape is the
$\theta$ angle of the twisted torus lattice. Within this
approximation the Casimir energy reads,
\begin{align}\label{telikiforforce}
&\mathcal{E}_c(s)=\frac{1}{(2\pi)^{D-1}}\frac{\pi^{\frac{D-1}{2}}}{\Gamma(s)}{\,}{\,}\times\\&
\notag
\Bigg{(}2{\,}\pi^{s-\frac{D-1}{2}}{\,}(R\sin\theta)^{-(s-\frac{D-1}{2})-\frac{1}{2}}{\,}(m^2)^{-\frac{1}{2}(s-\frac{D-1}{2})+\frac{1}{4}}\sum_{n=1}^{\infty}n^{s-\frac{D-1}{2}-\frac{1}{2}}K_{s-\frac{D-1}{2}-\frac{1}{2}}\big{(}2\pi
n{\,}mR\sin\theta\big{)}
\\&\notag+4\pi{\,}(m^2)^{-(s-\frac{D-1}{2})+1}(1-\cos^2\theta\sin^2\theta)^{-\frac{1}{2}}\Gamma(s-\frac{D-1}{2}-1)\sin \theta
R^2
\\&\notag+2^{3/2}{\,}(m^2)^{-(s-\frac{D-1}{2})+\frac{1}{4}}(1-\cos^2\theta\sin^2\theta)^{-\frac{1}{4}}{\,}\pi^{s-\frac{D-1}{2}}R^{\frac{3}{2}}\sin\theta\times
\\&\notag\sum_{n=1}^{\infty}n^{s-\frac{D-1}{2}-1}K_{s-\frac{D-1}{2}-\frac{1}{2}}\big{(}4\pi
n{\,}mR(1-\cos^2\theta\sin^2\theta)^{-\frac{1}{2}}\big{)}
\\&\notag
+2^{s-\frac{D-1}{2}+\frac{5}{2}}{\,}\pi^{s-\frac{D-1}{2}}R\sin\theta\sum_{n=1}^{\infty}n^{s-\frac{D-1}{2}-\frac{1}{2}}\cos(2n\pi\cos\theta\sin^2\theta)
\times
\\& \notag\sum_{d/n}d^{1-2(s-\frac{D-1}{2})}\Big{(}\frac{(1-\cos^2\theta\sin^2\theta)}{R^4\sin^2\theta}+\frac{4m^2}{R^2\sin^2\theta
d^2}\Big{)}^{-\frac{1}{2}{(s-\frac{D-1}{2})}+\frac{1}{4}}\times\\&\notag
K_{s-\frac{D-1}{2}-\frac{1}{2}}\Big{(}n\pi
R\sin\theta\sqrt{\frac{(1-\cos^2\theta\sin^2\theta)}{R^2}+\frac{4m^2}{d^2}}{\,}\Big{)}\Bigg{)}
\end{align}
Of course $D$=odd. We shall study the cases $D=3$ and $D=5$ but
the results are similar, really interesting and very permissible
for theories with compact extra dimensions.

\noindent In Figure \ref{energymassive1} we plot the Casimir
energy for $D=3$, $R=10^{-8}$ and $m=100$ and in Figure
\ref{energymassive2} for $D=5$. Also in Figure
\ref{energymassive3} we present the contour plot of the Casimir
energy as a function of $R$ and $\theta$. The lighter colors
correspond to larger values of the Casimir energy. Let us discuss
on these. As we can see the Casimir energy for both $D$=3 and
$D$=5 takes negative and positive values, with varying $\theta$.
The negative values of the Casimir energy is a very attractive
feature of theories with compact extra dimensions. Notice that
near the most studied case $\theta=\frac{\pi}{2}$ the Casimir
energy is negative. When the Casimir energy is negative (and more
and more negative as $R$ gets smaller) this leads to a shrinking
of the compact dimensions. This is true in the case the Casimir
energy contains inverse powers of $R$ and this is our case also.
Thus we see that near $\theta=\frac{\pi}{2}$, the Casimir energy
is negative. Behind this fact is the existence of an attractive
Casimir force. We now compute the Casimir force and we continue
soon this discussion. The computation of the Casimir force from
(\ref{telikiforforce}) is straightforward. The Casimir force
equals to,
\begin{equation}\label{casfor}
\mathcal{F}_c=-\frac{\partial\mathcal{E}_c}{\partial R}
\end{equation}
and using
\begin{equation}\label{paragogos}
\frac{\partial}{\partial
x}K_{\nu}(x{\,}z)=-\frac{1}{4}z{\,}K_{\nu-1}(x{\,}z)-\frac{1}{4}z{\,}K_{\nu+1}(x{\,}z)
\end{equation}
the Casimir force for the massive case reads,
\begin{align}\label{force}
&\mathcal{F}_c(s)=-\frac{1}{(2\pi)^{D-1}}\frac{\pi^{\frac{D-1}{2}}}{\Gamma(s)}{\,}{\,}\times\\&
\notag
\Bigg{(}2{\,}\pi^{s-\frac{D-1}{2}}{\,}(\sin\theta)^{\frac{1}{2}-(s-\frac{D-1}{2})}(-(s-\frac{D-1}{2})-\frac{1}{2})\times
\\&\notag R^{-(s-\frac{D-1}{2})-\frac{3}{2}}{\,}(m^2)^{-\frac{1}{2}(s-\frac{D-1}{2})+\frac{1}{4}}\sum_{n=1}^{\infty}n^{s-\frac{D-1}{2}-\frac{1}{2}}K_{s-\frac{D-1}{2}-\frac{1}{2}}\big{(}2\pi
n{\,}mR\sin\theta\big{)}\\& \notag
-\frac{1}{2}\pi^{s-\frac{D-1}{2}}{\,}(R\sin\theta)^{-(s-\frac{D-1}{2})-\frac{1}{2}}{\,}(m^2)^{-\frac{1}{2}(s-\frac{D-1}{2})+\frac{1}{4}}2\pi{\,}m\sin\theta\times\\&\notag
\sum_{n=1}^{\infty}n^{s-\frac{D-1}{2}+\frac{1}{2}}\Big{\{}K_{s-\frac{D-1}{2}+\frac{1}{2}}\big{(}2\pi
n{\,}mR\sin\theta\big{)}+K_{s-\frac{D-1}{2}-\frac{3}{2}}\big{(}2\pi
n{\,}mR\sin\theta\big{)}\Big{\}}
\\&\notag+8\pi{\,}(m^2)^{-(s-\frac{D-1}{2})+1}\Gamma(s-\frac{D-1}{2}-1)\sin \theta
R(1-\cos^2\theta\sin^2\theta)^{-\frac{1}{2}}
\\&\notag+2^{3/2}{\,}(m^2)^{-(s-\frac{D-1}{2})+\frac{1}{4}}(1-\cos^2\theta\sin^2\theta)^{-\frac{1}{4}}{\,}\pi^{s-\frac{D-1}{2}}\frac{3}{2}R^{\frac{1}{2}}\sin\theta\times \\&\notag \sum_{n=1}^{\infty}n^{s-\frac{D-1}{2}-1}K_{s-\frac{D-1}{2}-\frac{1}{2}}\big{(}4\pi
n{\,}mR(1-\cos^2\theta\sin^2\theta)^{-\frac{1}{2}}\big{)}
\\&\notag
-2^{3/2}{\,}(m^2)^{-(s-\frac{D-1}{2})+\frac{1}{4}}(1-\cos^2\theta\sin^2\theta)^{-\frac{1}{4}}{\,}\pi^{s-\frac{D-1}{2}}R^{\frac{3}{2}}{\,}\pi{\,}
m\sin\theta \times
\\&\notag \sum_{n=1}^{\infty}n^{s-\frac{D-1}{2}}\Big{\{}K_{s-\frac{D-1}{2}-\frac{3}{2}}\big{(}4\pi
n{\,}mR(1-\cos^2\theta\sin^2\theta)^{-\frac{1}{2}}\big{)}\\&\notag+K_{s-\frac{D-1}{2}+\frac{1}{2}}\big{(}4\pi
n{\,}mR(1-\cos^2\theta\sin^2\theta)^{-\frac{1}{2}}\big{)}\Big{\}}
\\&\notag
+2^{s-\frac{D-1}{2}+\frac{5}{2}}{\,}\pi^{s-\frac{D-1}{2}}\sin\theta\sum_{n=1}^{\infty}n^{s-\frac{D-1}{2}-\frac{1}{2}}\cos(2n\pi\cos\theta\sin^2\theta)\times
\\&\notag
\sum_{d/n}d^{1-2(s-\frac{D-1}{2})}\Big{(}\frac{(1-\cos^2\theta\sin^2\theta)}{R^4\sin^2\theta}+\frac{4m^2}{R^2\sin^2\theta
d^2}\Big{)}^{-\frac{1}{2}{(s-\frac{D-1}{2})}+\frac{1}{4}}\times
\\&\notag\Bigg{\{}K_{s-\frac{D-1}{2}-\frac{1}{2}}\Big{(}n\pi
R\sin\theta\sqrt{\frac{(1-\cos^2\theta\sin^2\theta)}{R^2}+\frac{4m^2}{d^2}}{\,}\Big{)}\\&\notag
-4\big{[}\frac{1}{4}-(s-\frac{D-1}{2})\big{]}K_{s-\frac{D-1}{2}-\frac{1}{2}}\Big{(}n\pi
R\sin\theta\sqrt{\frac{(1-\cos^2\theta\sin^2\theta)}{R^2}+\frac{4m^2}{d^2}}{\,}\Big{)}
\\&\notag -\frac{R}{2}\Big{(}\sqrt{\frac{(1-\cos^2\theta\sin^2\theta)}{R^2}+\frac{4m^2}{d^2}}\sin\theta {\,}\pi-\frac{\sin\theta {\,}\pi}{\sqrt{\frac{(1-\cos^2\theta\sin^2\theta)}{R^2}+\frac{4m^2}{d^2}}{\,}R^2}\Big{)}\times
\\&\notag \Big{[}K_{s-\frac{D-1}{2}+\frac{1}{2}}\Big{(}n\pi
R\sin\theta\sqrt{\frac{(1-\cos^2\theta\sin^2\theta)}{R^2}+\frac{4m^2}{d^2}}{\,}\Big{)}\\&\notag
+K_{s-\frac{D-1}{2}-\frac{3}{2}}\Big{(}n\pi
R\sin\theta\sqrt{\frac{(1-\cos^2\theta\sin^2\theta)}{R^2}+\frac{4m^2}{d^2}}{\,}\Big{)}\Big{]}\Bigg{\}}
\Bigg{)}
\end{align}

\begin{figure}[h]
\begin{center}
\includegraphics[scale=1.0]{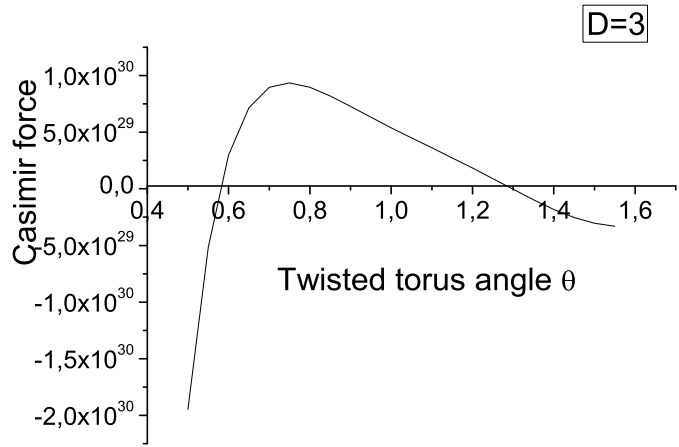}
\end{center}
\caption{The Casimir force $\mathcal{F}_{c}$ as a function of
$\theta$, the twisted torus angle, for $D=3$. (Massive case)}
\label{forcemassive3}
\end{figure}

\begin{figure}[h]
\begin{center}
\includegraphics[scale=.8]{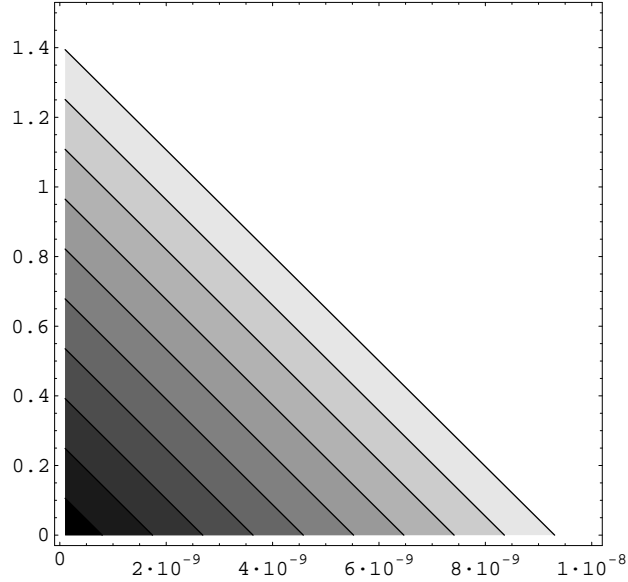}
\end{center}
\caption{Contour plot of the Casimir energy $\mathcal{E}_c$ as a
function of $\theta$ (vertical), the twisted torus angle, and $R$,
the compact radius (horizontal), for $D=3$. (Massive case)}
\label{contourforce}
\end{figure}

\noindent The Bessel series is converging really fast. Also in the
last term the argument of the Bessel function is $R$-independent.
Thus only a few terms of the last term give the dominating
contribution. During the process we ascertained that the other
Bessel sums are really fast convergent independently of the number
of terms we keep. In Figure \ref{forcemassive3} we plot the
Casimir force for $D=3$, $R=10^{-8}$ and $m=100$, and in Figure
\ref{contourforce} we present the contour plot of the force as a
function of $R$ and $\theta$.

\noindent As is seen from Fig. \ref{forcemassive3}, the Casimir
force as a function of $\theta$ behaves as the Casimir energy
does. As it can be seen the force and energy change sign as
$\theta$ varies. Also there exist $\theta$ values for which the
Casimir energy is completely zero. The most interesting cases are
those for which the energy and force are both negative. Indeed in
this case the internal space shrinks without limit. The Casimir
force is due to the non trivial topology of the compact space and
is responsible for the compact dimensions shrinking. It is
necessary for the compact space to be stabilized before it shrinks
to very small lengths. However this does not happen here, as it
can be easily checked. Indeed there is no stable minimum for the
Casimir energy as a function of $R$. We have not checked the case
when $R_1\neq R_2$ but this is out of the scopes of this paper.

\section{Massless Case}

In this section we calculate the scalar Casimir energy and Casimir
force for $M^D\times T^2_{\theta}$. It seems that this case is
very interesting phenomenologically since the calculations are
valid (that is singularity free) only for $D$=4 (!). This is
indeed surprising. Indeed in this case the regularized Casimir
energy for the twisted torus reads,
\begin{align}\label{vlad1}
&\mathcal{E}_c(s)=\frac{1}{(2\pi)^{D-1}}\int
\mathrm{d}^{D-1}p\sum^{\infty}_{n,{\,}n_1=-\infty}\Big{[}\sum_{k=1}^{D-1}p_k^2+an^2+b{\,}nn_1+cn_1^2\Big{]}^{-s}
\\&\notag -\frac{1}{(2\pi)^{D-1}}\int
\mathrm{d}^{D-1}p\Big{[}\sum_{k=1}^{D-1}p_k^2\Big{]}^{-s}.
\end{align}
and using (\ref{feynman}), we get,
\begin{align}\label{pordoulis213}
\mathcal{E}_c(s)=\frac{1}{(2\pi)^{D-1}}\pi^{\frac{D-1}{2}}\frac{\Gamma(s-\frac{D-1}{2})}{\Gamma(s)}\sum^{\infty
'}_{n,{\,}n_1=-\infty}\Big{[}an^2+b{\,}nn_1+cn_1^2\Big{]}^{\frac{D-1}{2}-s}.
\end{align}
Using the homogeneous Epstein zeta-like function,
\begin{equation}\label{inhomoge123}
E(s;a,b,c)=\sum^{\infty
'}_{n,{\,}n_1=-\infty}\Big{[}an^2+b{\,}nn_1+cn_1^2+m^2\Big{]}^{-s}
\end{equation}
and the Chowla-Selberg expansion that holds for it in this case
\cite{elizalde,elizaldekirsten},
\begin{align}\label{gone1}
&E(s;a,b,c)
=2{\,}\zeta(2s)a^{-s}+\frac{2^{2s}\sqrt{\pi}{\,}{\,}\Gamma(s-\frac{1}{2})\zeta(2s-1)}{\Gamma(s)\Delta^{s-\frac{1}{2}}}{\,}a^{s-1}
\\&\notag +\frac{2^{s+\frac{5}{2}}\pi^s{\,}}{\Gamma(s)\sqrt{a}{\,}\Delta^{\frac{s}{2}-\frac{1}{4}}}\sum_{n=1}^{\infty}n^{s-\frac{1}{2}}\sum_{d/n}d^{1-2s}\cos(\frac{n\pi b}{a})K_{s-\frac{1}{2}}\big{(}\frac{\pi
n\sqrt{\Delta}}{a}\big{)}
\end{align}
the Casimir energy in the massless case is written,
\begin{align}\label{teliki1}
&\mathcal{E}_c(s)=\frac{1}{(2\pi)^{D-1}}\frac{\pi^{\frac{D-1}{2}}}{\Gamma(s)}{\,}{\,}\times\\&
\notag
\Bigg{(}2{\,}\zeta(2s-D+1)a^{-(s-\frac{D-1}{2})}\Gamma(s-\frac{D-1}{2})\\&\notag
+\frac{2^{2s-D+1}\sqrt{\pi}{\,}{\,}\Gamma(s-\frac{D-1}{2}-\frac{1}{2})\zeta(2s-D)}{\Delta^{s-\frac{D-1}{2}-\frac{1}{2}}}{\,}a^{s-\frac{D-1}{2}-1}
\\&\notag+\frac{2^{s-\frac{D-1}{2}+\frac{5}{2}}{\,}\pi^{s-\frac{D-1}{2}}}{\Delta^{\frac{1}{2}(s-\frac{D-1}{2})-\frac{1}{4}}\sqrt{a}}\sum_{n=1}^{\infty}n^{s-\frac{D-1}{2}-\frac{1}{2}}\cos(\frac{n\pi
b}{a})
\sum_{d/n}d^{1-2(s-\frac{D-1}{2})}K_{s-\frac{D-1}{2}-\frac{1}{2}}\Big{(}\frac{\pi
n}{a}\sqrt{\Delta}{\,}\Big{)}\Bigg{)}
\end{align}
As we can see, the Casimir energy is free of divergences only when
$D$=4. This is because the combination of the gamma functions of
the first two terms gives always a divergent contribution. However
for $D$=4 the first term is zero since $\zeta(-4)=0$ (note that
$s=-\frac{1}{2}$). It is very surprising that something really
works solely for four dimensions and not for other dimensions. We
shall concentrate our study for the $R_1=R_2$ case. It is clear
that, as before, the only parameter that characterizes the shape
of the extra dimensions is the twisted torus angle, $\theta$.

\noindent In the case $R_1=R_2$ the Casimir energy (\ref{teliki1})
reads,
\begin{align}\label{casimirequal}
&\mathcal{E}_c(s)=\frac{1}{(2\pi)^{D-1}}\frac{\pi^{\frac{D-1}{2}}}{\Gamma(s)}{\,}{\,}\times\\&\notag
\Bigg{(}2{\,}\zeta(2s-D+1)(R^2\sin^2\theta)^{(s-\frac{D-1}{2})}\Gamma(s-\frac{D-1}{2})\\&\notag
+2^{2s-D+1}\sqrt{\pi}{\,}\Gamma(s-\frac{D-1}{2}-\frac{1}{2})\zeta(2s-D)\frac{R^{2s-D+1}}{(1-\sin^2\theta\cos^2\theta)^{s-\frac{D-1}{2}-\frac{1}{2}}}{\,}\sin\theta
\\&\notag+2^{s-\frac{D-1}{2}+\frac{5}{2}}{\,}\pi^{s-\frac{D-1}{2}}(\sin\theta)^{s-\frac{D-1}{2}+\frac{1}{2}}\frac{R^{2s-D+1}}{(1-\sin^2\theta\cos^2\theta)^{\frac{1}{2}(s-\frac{D-1}{2}-\frac{1}{2})}}\times
\\&\notag \sum_{n=1}^{\infty}n^{s-\frac{D-1}{2}-\frac{1}{2}}\cos(2\pi n\cos\theta\sin^2\theta)\sum_{d/n}d^{1-2(s-\frac{D-1}{2})}K_{s-\frac{D-1}{2}-\frac{1}{2}}\Big{(}\pi n(1-\sin^2\theta\cos^2\theta)^{\frac{1}{2}}\sin\theta\Big{)}\Bigg{)}
\end{align}

\begin{figure}[h]
\begin{center}
\includegraphics[scale=1.0]{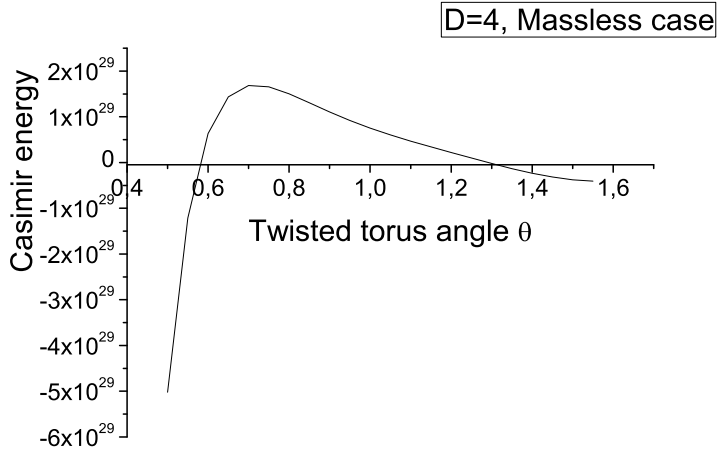}
\end{center}
\caption{The Casimir energy $\mathcal{E}_{c}$, for the massless
case, as a function of $\theta$, the twisted torus angle, with
$D=4$.} \label{energymassless}
\end{figure}

\noindent Thus the Casimir force reads,
\begin{align}\label{casimirforce}
&\mathcal{F}_c(s)=-\frac{1}{(2\pi)^{D-1}}\frac{\pi^{\frac{D-1}{2}}}{\Gamma(s)}{\,}{\,}\times\\&\notag
\Bigg{(}2{\,}\zeta(2s-D+1)2(s-\frac{D-1}{2})R^{2(s-\frac{D-1}{2})-1}{\,}(\sin^2\theta)^{(s-\frac{D-1}{2})}{\,}\Gamma(s-\frac{D-1}{2})\\&\notag
+2^{2s-D+1}\sqrt{\pi}{\,}\Gamma(s-\frac{D-1}{2}-\frac{1}{2}){\,}\zeta(2s-D)\Big{(}2s-D+1\Big{)}{\,}\frac{R^{2s-D}}{(1-\sin^2\theta\cos^2\theta)^{s-\frac{D-1}{2}-\frac{1}{2}}}\sin\theta
\\&\notag+2^{s-\frac{D-1}{2}+\frac{5}{2}}{\,}\pi^{s-\frac{D-1}{2}}\Big{(}2s-D+1\Big{)}(\sin\theta)^{s-\frac{D-1}{2}+\frac{1}{2}}{\,}\frac{R^{2s-D}}{(1-\sin^2\theta\cos^2\theta)^{\frac{1}{2}(s-\frac{D-1}{2}-\frac{1}{2})}}\times
\\&\notag \sum_{n=1}^{\infty}n^{s-\frac{D-1}{2}-\frac{1}{2}}\cos(2\pi n\cos\theta\sin^2\theta)\sum_{d/n}d^{1-2(s-\frac{D-1}{2})}K_{s-\frac{D-1}{2}-\frac{1}{2}}\Big{(}\pi n{(1-\sin^2\theta\cos^2\theta)^{\frac{1}{2}}}\sin\theta\Big{)}\Bigg{)}
\end{align}

\begin{figure}[h]
\begin{center}
\includegraphics[scale=1.0]{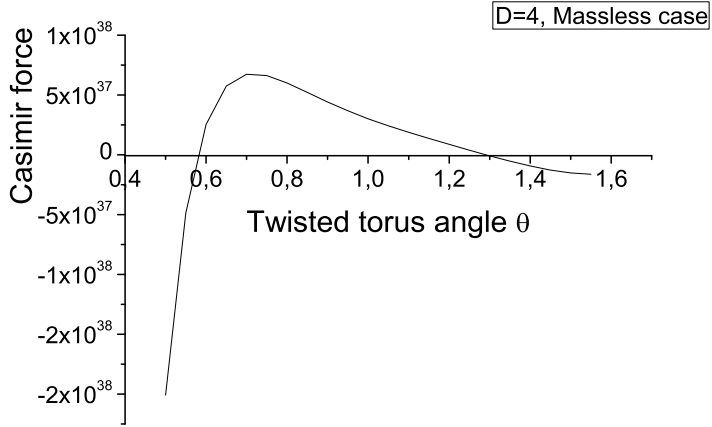}
\end{center}
\caption{The Casimir force $\mathcal{F}_{c}$, for the massless
case, as a function of $\theta$, the twisted torus angle, for
$D=4$.} \label{forcemassless}
\end{figure}

\begin{figure}[h]
\begin{center}
\includegraphics[scale=.8]{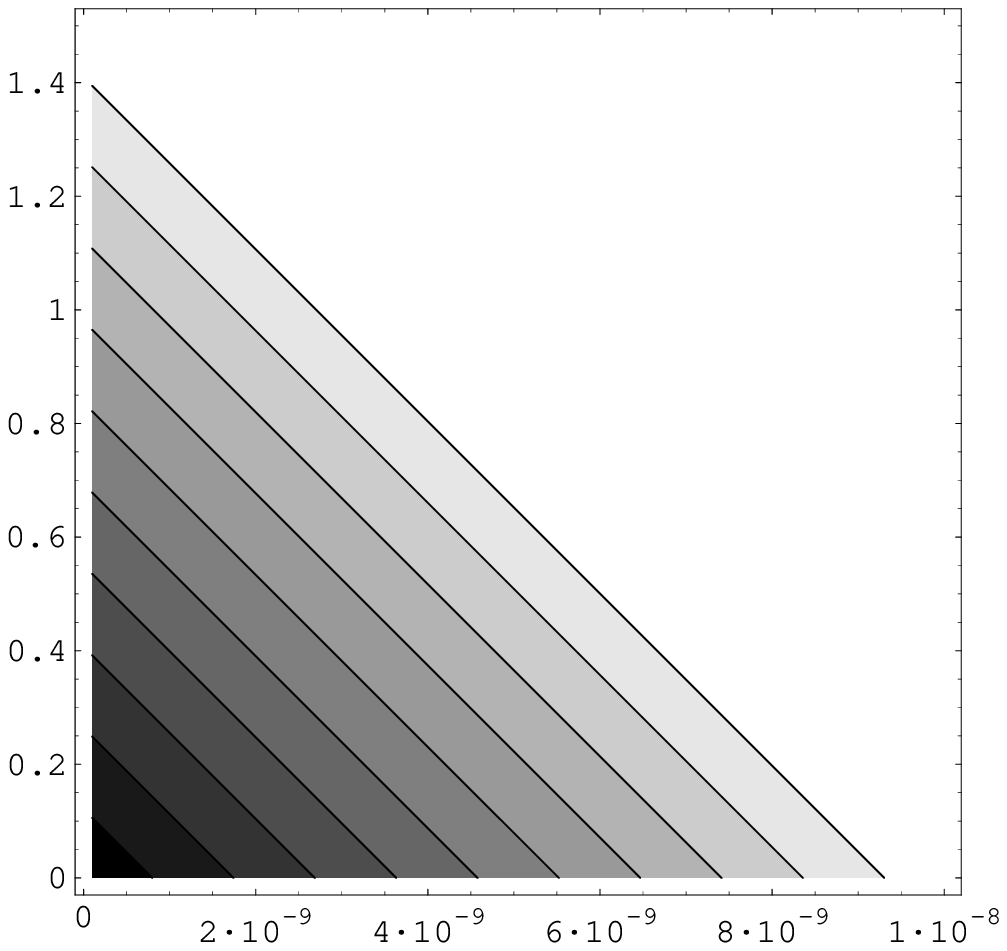}
\end{center}
\caption{Contour plot of the Casimir energy $\mathcal{E}_c$ as a
function of $\theta$ (vertical), the twisted torus angle, and the
compact radius $R$ (horizontal), for $D=4$.}
\label{contourenergymassless}
\end{figure}

\begin{figure}[h]
\begin{center}
\includegraphics[scale=.8]{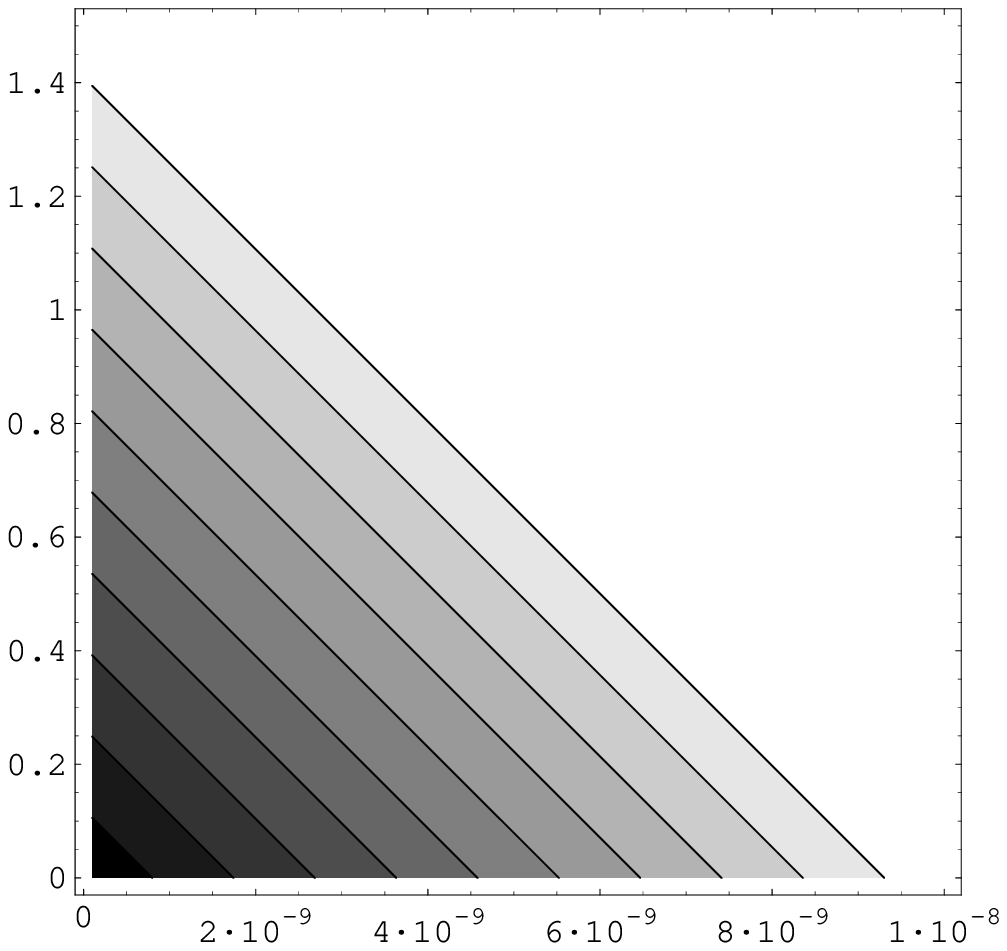}
\end{center}
\caption{Contour plot of the Casimir force $\mathcal{F}_c$ as a
function of $\theta$ (vertical), the twisted torus angle, and the
compact radius $R$ (horizontal), for $D=4$.}
\label{contourforcemassless}
\end{figure}

\noindent In figures \ref{energymassless} and \ref{forcemassless}
we plot the Casimir energy for $D=4$ as a function of $\theta$.
Also in figures \ref{contourenergymassless} and
\ref{contourforcemassless} we present the contour plots of the
Casimir energy and Casimir force respectively.

\noindent The analysis of this case is very similar with the
previous analysis. Both the Casimir energy and Casimir force
change sign and become positive and negative as $\theta$ varies.
As we mentioned previously the most viable case (at least
phenomenologically) arises when the Casimir energy is negative and
also goes to minus infinity as $R$ decreases (see the contour
plot). This is indeed our case and the Casimir energy is negative
for some values of $\theta$. For the same values of $\theta$ the
force behaves exactly in the same way as the energy does. We can
see that both the force and the energy can take very small
positive and negative values and also there exist values of
$\theta$ for which both of them are zero.

\noindent Another necessary issue to be studied is the stability
of the compact dimensions. The shrinking of the compact spaces
occurs in this setup but the stability does not. The Casimir
energy does not have a stable minimum. We have not studied the
$R_1\neq R_2$ case, which would be more rich in phenomenology but
is out of the scope of this article. In the last case the
shadowing effects would appear and this would make it an
independent exercise, that is to find how the Casimir energy
behaves for various dimensions and for the twisted torus for
various $\theta$. We shall discuss on these issues in the next
section.

\section{Conclusions}

\noindent We have studied how the shape of a twisted toroidal
compact space can affect the Casimir energy and Casimir force of a
scalar field quantized on spacetime of the form $M^D\times
T^2_{\theta}$. Both the massive and massless case were taken into
account. The Casimir energy was calculated for general $R_1$,
$R_2$ and $\theta$. We specified our study to the case $R_1=R_2$.
The main interest is to see how $\theta$ modifies the Casimir
energy and force, so the shape is represented by this parameter.
The compact dimensions radius was taken to be $\sim 10^{-8}$,
which is compatible with the ADD models predictions and the
current experimental bounds.

\noindent It was found that for the massive case (we took $mR\ll
1$, but this does not modify the results), the Casimir energy
contains infinities for $D$=even. Thus the Casimir energy for
massive fields cannot be computed for our spacetime in a
consistent way. One could naively say that the singularities could
be regularized in some way, however the fact that the
singularities are $R$-dependent and also the fact that for odd
dimensional spaces the Casimir energy is regular, makes us sure
that no consistent result for $D$=even holds, at least for the
topology of the compact twisted torus. Also the singularities
remain when the Casimir force is computed. Thus unfortunately no
bulk massive scalar field Casimir energy and force cannot be
computed for our spacetime, with this compact extra space. We
applied the study for $D=3$ and $D=5$ (although this is a
mathematical exercise it is interesting to compare it to the
massless case where spacetime dimensions $D=4$ is the only allowed
case!). We found similar results both for the two cases.
Particularly we found that the energy and force as a function of
$\theta$ become positive and negative as $\theta$ varies. Also
they can become zero for some $\theta$. Of course the most
interesting cases are when the energy and force is negative as we
discussed in the previous sections.

\noindent In contrast to the massive case, the massless scalar
Casimir energy can be computed consistently (that is free of
singularities) only for $D=4$! This is indeed very striking since
the $D=4$ spacetime dimensionality is very peculiar topologically
when quantum field theory calculations are performed. We used the
same values of $R$ as before and studied how the energy and force
behave as a function of $\theta$. We found that the behavior is
similar to the massive case, that is, both the energy and force
become positive and negative, also for some $\theta$ both take
very small values and additionally become zero.

\noindent Furthermore we briefly addressed the issue of the
stability of the compact space. We saw that the stabilization does
not occur for both massive and massless case. This holds of course
for the case $R_1=R_2$. We have not studied the $R_1\neq R_2$ case
because this case deserves a separate study due to the reach
phenomenology that exists for this case. Indeed as found by K.
Dienes in \cite{dienes}, in this case shadowing effects can take
place. Shadowing makes difficult the experimental detection of the
exact number and the true geometry of the compact dimensions
\cite{dienes}. Also interesting phenomena arise when $R_1/R_2$ is
a rational number. We hope to address this case soon.

\noindent Concerning the shadowing effects, when non trivial
identifications hold in the compact extra dimensional space,
similar to shadowing effect phenomena hold. In reference
\cite{sol} it was found that the exponential correction to the
Newton law force range varies, as the parameters of the
identifications change. Thus it is true that we would not be able
to be sure on what the number and the geometry of the extra
dimensions are.

\noindent Before ending we must note the significance of the
Casimir energy of bulk fields to cosmology and specifically to
dark energy which was addressed in \cite{levin} and also in
\cite{Miltonnew}. In \cite{levin} the extra compact space had
toroidal topology. This corresponds to $\theta=\frac{\pi}{2}$ in
our case. It would be interesting to find the cosmological
implications of a non-trivial $\theta$ to the dark energy.


\newpage

\begin{thebibliography}{99}



\bibitem{Casimir} H. Casimir, Proc. Kon. Nederl. Akad. Wet. 51 793
(1948)

\bibitem{dienes} K. R. Dienes, Phys. Rev. Lett. 88, 011601 (2002);
K. R. Dienes, A. Mafi, Phys. Rev. Lett. 88, 111602 (2002); K. R.
Dienes, arXiv:hep-ph/0211211

\bibitem{elizaldeoriginal} E. Elizalde, J. Math. Phys. 35, 6100
(1994); E. Elizalde, Commun. Math. Phys. 198, 83 (1998)

\bibitem{elizaldekirsten} K. Kirsten, E. Elizalde, Phys. Lett.
B365, 72 (1996)
\bibitem{levin}B. Greene, J. Levin, JHEP07011, 096 (2007)

\bibitem{Miltonnew}K. A. Milton, Grav. Cosm. 9, 66 (2003)

\bibitem{kanti}R. Hofmann, P. Kanti, M. Pospelov, Phys. Rev. D63,
124020 (2001)

\bibitem{sundrum}Mariana Frank, Nasser Saad, Ismail Turan,
arXiv:0807.0443

\bibitem{perivolaropoulos} L. Perivolaropoulos, Phys. Rev. D77,
107301 (2008)


\bibitem{Bordagreview} M. Bordag, U. Mohideen, V. M. Mostepanenko,
Phys. Rep. 353, 1 (2001)

\bibitem{elizalde} E. Elizalde, "Ten physical
applications of spectral zeta functions", Springer (1995); E.
Elizalde, S.~D.~Odintsov, A. Romeo, A. A. Bytsenko, "Zeta
regularization techniques and applications", World Scientific
(1994)

\bibitem{kirstenbook} Klaus Kirsten, Spectral Functions in Mathematics and
Physics,  Chapman Hall/CRC (2001)

\bibitem{kirsten12}K. Kirsten, Generalized multidimensional Epstein zeta functions, J. Math. Phys. 35, 459-470 (1994)

\bibitem{kirsten14}K. Kirsten,  Inhomogeneous multidimensional Epstein zeta functions, J. Math. Phys. 32, 3008-3014 (1991)

\bibitem{chodos} T. Appelquist, A. Chodos, Phys. Rev. Lett. 50 141
(1983)

\bibitem{gradsteyn} I.S. Gradshteyn and I.M. Ryzhik, Table of Integrals Series and Products
(Academic Press, 1965)

\bibitem{sol} V. K. Oikonomou, Class. Quant. Grav. 25, 195020 (2008)

\end{thebibliography}
\end{document}